\documentclass[12pt]{article}
\usepackage{amsmath}
\usepackage{graphicx}
\usepackage{dcolumn}
\usepackage{bm}
\usepackage{amssymb}
\usepackage{latexsym}

\begin{document}

\title{On the Contributions from Dilatonic Strings to the Flat Behaviour of the Rotational
Curves in Galaxies}

\author{M. Leineker Costa$^1$, A. L. N. Oliveira$^2$ and M. E. X.
Guimar\~aes$^3$ \\
\mbox{\small{1. Instituto de F\'{\i}sica, Universidade de
Bras\'{\i}lia, Bras\'{\i}lia-DF, Brazil}} \\
\mbox{\small{2. Departamento de F\'{\i}sica, Universidade Estadual
de Londrina, Londrina-PR, Brazil}} \\
\mbox{\small{3. Departamento de Matem\'atica, Universidade de
Bras\'{\i}lia, Bras\'{\i}lia-DF, Brazil}}}

\maketitle

\begin{abstract}
We analyse the flat behaviour of the rotational curves in some
galaxies in the framework of a dilatonic, current-carrying string.
We determine the expression of the tangential velocity of test
objects following a stable circular equatorial orbit in this
spacetime.
\end{abstract}



\section{Introduction:}

The measurements of rotation curves in some galaxies show that the
coplanar orbital motion of gas in the outer parts of the galaxies
keeps a constant velocity up to several luminous radii
\cite{matos, rubin, persic, persic2}. The most accepted
explanation for this effect is that there exists a spherical halo
of dark matter which surrounds the galaxy and account for the
missing mass needed to produce the flat behavior of the rotational
curves.

In this work, we would like to analyse this effect in the
framework of a dilatonic, current-carrying cosmic string. In a
number of papers [5-11], a current has been included in the
internal structure of the cosmic string. The most noticeable
consequence of a current-like effect is to modify the internal
dynamics of cosmic strings in such a way that new states are
reachable. Indeed, the breaking of the Lorentz boost invariance
along the worldsheet allows rotating equilibrium configurations,
called vortons, which, if they are stable, can overclose the
universe, thereby leading to a catastrophe for the theory that
predicts them~\cite{vortons}. Finally, inclusion of such an
internal structure could drastically change the predictions of a
cosmic string model~\cite{rdp} in the microwave background
anisotropies~\cite{CMB}. In \cite{vanessa}, it is shown that the
long-range effect on a cosmologically relevant network of strings
is vanishing on average, but that vorton-like states can be
reached by microscopically small loops. Here, we would like
address ourselves to a well-posed problem at the galactic scale.

In what follows, after having set the relevant gravitational
theory and notations, we derive the geometry of an electrically
charged cosmic string in the dilatonic theory of gravity in Sec.
2. Then, in the Sec. 3, we compute the tangential velocity for
test particles imposing that its magnitude is independent of the
radius. In doing this, we find a constraint equation for the
metric coefficients of the dilatonic cosmic string. Unfortunately,
as we will see later, we also find that the tangential velocity
cannot be explained by a single string of the kind proposed in our
model. Instead, in order to be compatible with the observed
magnitude, one must have a bundle of $N \sim 10^{5}$ strings
seeding a galaxy. With such a density, a cosmic string network
would be dominating the universe, and its dynamics would be
completely different. Section 4 summarizes our findings and
discusses the relevant conclusions.

\section{Exterior Solution for a Timelike Current-Carrying String:}

\subsection{The Model:}

In this section we will mainly review the obtention of the
gravitational field generated by a string carrying a current of
timelike-type as presented in the Refs. \cite{oliveira}.

We will concentrate our attention to superconducting vortex
configurations which arise from the spontaneous breaking of the
symmetry $U(1) \times U_{em}(1)$. Therefore, the action for the
matter fields will be composed by two pairs of coupled complex
scalar and gauge fields $(\varphi, B_{\mu})$ and $(\sigma,
A_{\mu})$. Also, for technical purposes, it is preferable to work
in the so-called Einstein (or conformal) frame, in which the
scalar and tensor degrees of freedom do not mix:
\begin{eqnarray}
{\cal{S}}&=& \frac{1}{16 \pi G_*} \int d^4x \sqrt{-g} [R-2\partial _{\mu} \phi  \partial^{\mu} \phi ] \nonumber \\
&+& \int d^4x \sqrt{-g} [ -\frac{1}{2}\Omega ^2 (\phi) ((D_{\mu}\varphi) ^{*} D^{\mu} \varphi + (D_{\mu} \sigma) ^{*} D^{\mu} \sigma) \nonumber \\
&-& \frac{1}{16 \pi}  ( F_{\mu \nu} F^{\mu \nu} + H_{\mu \nu} H^{\mu \nu}) - \Omega ^2(\phi)  V(|\varphi|,|\sigma|) ] ,
\end{eqnarray}
where $F_{\mu \nu} = \partial{\mu} A_{\nu} - \partial _{\nu}
A_{\mu}$ , $H_{\mu \nu }= \partial{\mu} B_{\nu} - \partial _{\nu}
B_{\mu}$ and the potential is suitably chosen in order that the
pair $(\varphi, B_{\mu})$ breaks one symmetry $U(1)$ in vacuum
(giving rise to the vortex configuration) and the second pair
$(\sigma, A_{\mu})$ breaks the symmetry $U_{em}(1)$ in the core of
the vortex (giving rise to the superconducting properties): $
V(|\varphi|,|\sigma|) = \frac{\lambda _{\varphi}}{8} (|\varphi|^2-
\eta ^2)^2 + f(|\varphi|^2 - \eta ^2) |\sigma|^2 + \frac{\lambda
_{\sigma}}{4} |\sigma|^4 + \frac{m_{\sigma}^2}{2} |\sigma|^2 \,\,
.$

In what follows, we will write the general static metric with
cylindrical symmetry corresponding to the electric case in the
form
\begin{equation}
\label{metrica1}
ds^2 = - e^{2\psi}dt^2 +
 e^{2(\gamma - \psi)}(dr^2 + dz^2) + \beta^2 e^{-2\psi}d\theta^2
\end{equation}
where $\psi, \gamma, \beta$ are functions of $r$ only.


In order to solve the equations we will divide the space in two
regions: the exterior region, $r \geq r_0$, in which only the
electric component of the Maxwell tensor contributes to the
energy-momentum tensor and the internal region, $0 \leq r < r_0$,
where all matter fields survive. $r_0$ is the string thickness.

\subsection{Form of the Line Element for the Exterior Region:}

Due to the specific properties of the Maxwell tensor
\begin{equation}
T^{\mu}_{\mu} =0 \hspace{1cm} \mbox{and} \hspace{1cm} T^{\alpha}_{\nu}T^{\mu}_{\alpha} = \frac{1}{4}
(T^{\alpha \beta} T_{\alpha \beta}) \delta^{\mu}_{\nu}
\end{equation}
we can find the metric through some  algebraic relations called
Rainich algebra \cite{louis, melvin}, which, for the  electric
case, have the form\footnote{In the scalar-tensor theories, these
relations are modified by a term which depend on the dilaton
\cite{fer, oliveira}.}:
\[
R^t_t=-R^{\theta}_{\theta} \hspace{0.3cm} R^t_t=R^r_r-2g^{rr}
{\phi ^\prime}^2 \hspace {0.3cm} R^{\theta}_{\theta}=R^z_z \,\, .
\]

Therefore, the exterior metric for a timelike current-carrying string is:
\begin{eqnarray}
\label{extele}  ds^2_E & = &
\left(\frac{r}{r_0}\right)^{2m^2-2n}W^2(r) (dr^2+dz^2)
\nonumber \\
&+& \left(\frac{r}{r_0}\right)^{-2n}W^2(r) B^2r^2d\theta^2
\nonumber \\
&-&
\left(\frac{r}{r_0}\right)^{2n}\frac{1}{W^2(r)}dt^2
\end{eqnarray}
where
\begin{equation}
\label{w}
W(r)\equiv \frac{\left(\frac{r}{r_0}\right)^{2n}+k}{1+k} \, .
\end{equation}
The constants $m , n, k, B$ will be determined after the inclusion
of the matter fields.

\section{Stable Circular Geodesics Around the Cosmic String:}

In this section we will derive the geodesic equations in the
equatorial plane ($\dot{z}=0$), where dot stands for ``derivative
with respect to the proper time $\tau$". First of all, let us
re-write metric (4) in a more compact way:
\begin{equation}
\label{comp}
 ds^2 = A(r)[dr^2 + dz^2] + B(r) d\theta^2 - C(r) dt^2 \,\, ,
\end{equation}
with
\begin{eqnarray}
A(r) & = & \left(\frac{r}{r_0}\right)^{2m^2-2n}W^2(r) \,\, , \nonumber \\
B(r) & = & \left(\frac{r}{r_0}\right)^{-2n}W^2(r)\beta^2(r) \,\, , \nonumber \\
C(r) & = & \left(\frac{r}{r_0}\right)^{2n}W^{-2}(r) \,\, .
\nonumber
\end{eqnarray}

The Lagrangian for a test particle moving on this spacetime is
given by:
\begin{equation}
2{\cal{L}} = A(r)[\dot{r}^2 + \dot{z}^2] + B(r)\dot{\theta}^2
-C(r)\dot{t}^2
\end{equation}
and the associated canonical momenta, $p_{\alpha}=
\frac{\partial{\cal{L}}}{\partial\dot{x}^{\alpha}}$, are:
\begin{eqnarray}
p_t & = & - E = -C(r)\dot{t} \,\, , \nonumber \\
p_{\theta} & = & L = B(r)\dot{\theta} \,\, , \nonumber \\
p_r & = & A(r) \dot{r} \,\, , \nonumber \\
p_z & = & A(r)\dot{z} \,\, .
\end{eqnarray}
Because of the symmetries of this particular spacetime, the
quantities $E$ and $L$ are constants for each geodesic and,
because this spacetime is static, the Hamiltonian, ${\cal{H}} =
p_{\alpha}\dot{x}^{\alpha} - {\cal{L}}$, is also a constant.
Combining this information with the restriction of a motion in an
equatorial plane, we arrive to the following equation for the
radial geodesic:
\begin{equation}
\dot{r}^2 - A^{-1}[E\dot{t} - L\dot{\theta} - 1]=0 \,\, .
\end{equation}

In this work, we will concentrate on stable circular motion.
Therefore, we have to satisfy three conditions simultaneously.
Namely:
\begin{itemize}
\item $\dot{r} =0$ \,\, ; \item $\frac{\partial{V(r)}}{\partial r}
= 0$, where $V(r) = - A^{-1}[E\dot{t} - L\dot{\theta} - 1]$  \,\,
; \item $\frac{\partial^2{V(r)}}{\partial r^2}\mid_{ext} \,  > 0$,
in order to have a minimum \,\, .
\end{itemize}
Consequently, we have:
\begin{eqnarray}
E\dot{t} - L\dot{\theta} - 1 & = & 0 \nonumber \\
\frac{\partial}{\partial r}\left\{A^{-1}[E\dot{t} - L\dot{\theta}
- 1]\right\} & = & 0 \,\, .
\end{eqnarray}
Expressing $\dot{t}$ and $\dot{\theta}$ in terms of the constant
quantities $E$ and $L$, respectively, we can get their expressions
\footnote{It is remarkable that our conclusions are independent on
the metric coefficient $A$.}:
\begin{eqnarray}
E & = & C\sqrt{\frac{B'}{B'C-BC'}} \,\, , \nonumber \\
L & = & B \sqrt{\frac{C'}{B'C-BC'}} \,\, .
\end{eqnarray}
where prime means ``derivative with respect to the coordinate
$r$".

Recalling that the angular velocity of a test particle moving in a
circular motion in an orbital plane is $\Omega =
\frac{d\theta}{dt} = \frac{\dot{\theta}}{\dot{t}}$, we have:
\begin{equation}
\Omega = \sqrt{\frac{C'}{B'}} \,\, .
\end{equation}
We are now in a position to compute the tangential velocity of the
moving particles in a circular orbit in the equatorial plane. From
now on, we will follow the prescription established by
Chandrasekhar in \cite{chandra}. Let us re-express the metric (6)
in terms of the proper time $\tau$, as $d\tau^2 = - ds^2$:
\begin{equation}
d\tau^2 = C(r)dt^2 \left\{ 1 -
\frac{A}{C}\left[\left(\frac{dr}{dt}\right)^2 +
\left(\frac{dz}{dt}\right)^2\right]
-\frac{B}{C}\left(\frac{d\theta}{dt}\right)^2\right\} \, ,
\end{equation}
and comparing with the expression $ 1 = C(r)(u^0)^2[1 - v^2] \,\,
,$ where $u^0 = dt/d\tau$, we can easily obtain the spatial
velocity $v^2$:
\begin{equation}
v^2 = (v^{(r)})^2 + (v^{(z)})^2 + (v^{(\theta)})^2 \,\, ,
\end{equation}
from which we can obtain  all the components of the spatial
velocities. However, we are particularly interested in the
tangential component $v^{(\theta)}$:
\begin{equation}
v^{(\theta)} =  \sqrt{\frac{B}{C}}\left(\frac{d\theta}{dt}\right)
= \sqrt{\frac{B}{C}}\Omega \,\, .
\end{equation}
In order to have stable circular orbits, the tangential velocity
$v^{(\theta)}$ must be constant at different radii at the
equatorial plane. Therefore, we can impose:
\begin{equation}
v^{(\theta)} =\sqrt{\frac{BC'}{B'C}} = v^{(\theta)}_c =
\mbox{const.}
\end{equation}
At this point we would like to notice the following theorem
\cite{matos}:

{\bf{\underline{Theorem:}}} \hspace{0.3cm}{\it The tangential
velocity of circular stable equatorial orbits is constant iff the
coefficient metrics are related as}\footnote{We point out to the
reader that, for the sake of clarity, we have changed the notation
with respect to the original paper \cite{matos}. The Theorem, of
course, is unchangeable.}
\begin{equation}
\label{theorem}
 C^{1/2} = \left(\frac{r}{r_0}\right)^l \,\, ,
\end{equation}
{\it provided $l = \frac{v^{(\theta)}_c}{1+v^{(\theta)}_c}$.}

This theorem implies that the line element in the equatorial plane
must be\footnote{Here, again, we have adapted the notation for the
sake of clarity.}:
\begin{equation}
\label{form} ds^2 = -\left(\frac{r}{r_0}\right)^{2l}dt^2 +
\left(\frac{r}{r_0}\right)^{-2l}[\left(\frac{r}{r_0}\right)^{2m^2}dr^2
+ B^2r^2d\theta^2] \,\, .
\end{equation}
This form clearly is not asymptotically flat and neither describes
a spacetime corresponding to a central black hole. Therefore, we
can infer that it describes solely the region where the tangential
velocity of the test particles is constant, being probably joined
in the interior and exterior regions with other metrics, suitably
chosen in order to ensure regularity in the asymptotic limits.

Let us notice, however, that this metric has the form which has
been found in \cite{oliveira}, after identifying $l$ with the
appropriate constant parameters which depend on the microscopic
details of the model. The calculations are straightforward but
lengthy. We will skip here the details and provide directly the
results. For the details of these calculations, we refer the
reader to Refs. \cite{oliveira}. For this particular
configuration, consisting of an electrically charged dilatonic
string, we have:
\begin{equation}
\label{l}
 l= 2G_0\alpha(\phi_0) [U + T + I^2] \,\, ,
\end{equation}
where $U,T$ and $I^2$ are the energy per unit length, the tension
per unit length and the current of the string, respectively.
$\alpha(\phi_0)$ measures the coupling of the dilaton to the
matter fields. For cosmic strings formed at GUT scales, $G_0 [U +
T + I^2] \sim 3 \times 10^{-6}$, and for a coupling
$\alpha(\phi_0)$ which is compatible with present experimental
data \cite{damour}, $\alpha(\phi_0) < 10^{-3}$, the parameter $l$
(and, thus, the tangential velocity $v^{(\theta)}_c$) seems to be
too small. The observed magnitude of the tangential velocity
\cite{rubin, persic, persic2} being $v^{(\theta)}_c > 3 \times
10^{-4}$ cannot be explained by a single dilatonic
current-carrying string in this case. As argued by Lee in the Ref.
\cite{lee}, if a bundle of $N$ cosmic strings formed at GUT scales
seeded one galaxy, then the total magnitude of the tangential
velocity would be $Nv^{(\theta)}_c$. In our case, to be compatible
with the astronomical observations, one must have a bundle of $N
\sim 10^{5}$ strings seeding a galaxy. With such a density, a
cosmic string network would be dominating the universe, and its
dynamics would be completely different.  The only situation where
such a high number of strings could be  possible is at much lower
energy scales (electroweak scale, say) but then of course the
energy scale is far too low to have any relevance for structure
formation.

If, on the other hand, one supposes that one single string could
explain the observed values of the tangential velocity, one should
therefore impose that such a string is formed at Planck scales,
which at the moment seems not quite realistic conclusion as well.

\section{Conclusions:}

The measurements of rotation curves in galaxies show that the
coplanar orbital motion of gas in the outer parts of the galaxies
keeps a constant velocity up to several luminous radii. The most
accepted explanation for this effect is that there exists a
spherical halo of dark matter which surrounds the galaxy and
account for the missing mass needed to produce the flat behavior
of the rotational curves.

In the Ref. \cite{matos}, it has been shown that in a static,
axially symmetric spacetime, a sufficient and necessary condition
in order to have a flat behavior for the rotational curves in
galaxies is that the metric assumes the form (\ref{form}). This
form clearly is not asymptotically flat and neither describes a
spacetime corresponding to a central black hole. Therefore, we can
infer that it describes solely the region where the tangential
velocity of the test particles is constant, being probably joined
in the interior and exterior regions with other metrics, suitably
chosen in order to ensure regularity in the asymptotic limits.

In previous papers \cite{oliveira}, we have found a metric
corresponding to the spacetime generated by an electrically
charged dilatonic string which possesses the form (\ref{form}),
with appropriate parameters which are related to the microscopic
details of the model (\ref{l}). The observed magnitude of the
tangential velocity cannot be explained by a single dilatonic
current-carrying string in this case. However, if we consider that
a bundle of $N$ cosmic strings formed at GUT scales seeded one
galaxy, then the total magnitude of the tangential velocity would
be $Nv^{(\theta)}_c$. In our case, to be compatible with the
astronomical observations, one must have a bundle of $N \sim
10^{5}$ strings seeding one galaxy! The only situation where such
a high number of strings could be  possible is at much lower
energy scales (electroweak scale, say) but then the energy scale
is far too low to have any relevance for structure formation.

\section*{Acknowledgements:}

The authors would like to thank Prof. E. R. Bezerra de Mello for
very interesting comments and suggestions. M. Leineker Costa would
like to thank CAPES for a grant. A. L. Naves de Oliveira and M. E.
X. Guimar\~aes would like to thank CNPq for a support. M. E. X.
Guimar\~aes would like to thanks the High Energy Sector of the
Abdus Salam International Center for Theoretical Physics for
hospitality during the preparation of this work.


\begin{thebibliography}{99}

\bibitem{matos} T. Matos, D. N\'u\~nez, F. Siddartha Guzm\'an and
E. Ram\'{\i}rez, Gen. Rel. Grav. {\bf 34}, 283 (2002).

\bibitem{rubin} V. C. Rubin, N. Thonnard and W. K. Ford, Ap. J.
{\bf 225}, L107 (1978); Ap. J. {\bf 238}, 471 (1980).

\bibitem{persic}M. Persic and P. Salucci, Ap.J. Suppl. Ser. {\bf
99}, 501 (1995).

\bibitem{persic2}M. Persic, P. Salucci and F. Stel, M. N. R. A. S. {\bf 281},
27 (1996).

\bibitem{peter}  P.~Peter, Class. Quantum Grav. {\bf 11}, 131 (1994).

\bibitem{witten}  E.~Witten, Nucl. Phys. {\bf B249}, 557, (1985).

\bibitem{formal} B.~Carter, Phys. Lett. B {\bf 224}, 61 (1989); {\bf
228}, 466 (1989); Nucl. Phys. B {\bf 412}, 345 (1994); see also
B.~Carter, P.~Peter, and A.~Gangui, Phys. Rev. D {\bf 55}, 4647
(1997).

\bibitem{neutral} P.~Peter, Phys. Rev. D {\bf 45}, 1091 (1992); {\bf 46},
3335 (1992).

\bibitem{fer} C.~N.~Ferreira, M.~E.~X.~Guimar\~aes, and
J.~A.~Helay\"el-Neto, Nucl. Phys. {\bf B581}, 165 (2000).

\bibitem{oliveira} A.~L.~N.~Oliveira and M.~E.~X~Guimar\~aes,
Phys. Lett. A {\bf 311}, 474 (2003); A.~L.~N.~Oliveira and
M.~E.~X~Guimar\~aes, Phys. Rev. D {\bf 67}, 123514 (2003).

\bibitem{van} M. C. B. Abdalla, A. A. Bytsenko and M. E. X. Guimar\~aes,
Mod. Phys. Lett. A {\bf 19}, 2445 (2004); V. C. Andrade, A. L.
Naves de Oliveira and M. E. X. Guimar\~aes, Proc. Sci.
WC2004(2004)034.

\bibitem{vortons} R.~L.~Davis, Phys. Rev. D {\bf 38}, 3722 (1988); R.~L.~Davis
and E.~P.~S.~Shellard, Nucl. Phys. {\bf B323}, 209 (1989);
B.~Carter, Ann. N.Y. Acad.Sci. {\bf 647}, 758 (1991); B.~Carter,
Phys. Lett. B {\bf 238}, 166 (1990); in {\it Proceedings of the
XXXth Rencontres de Moriond, Villard--sur--Ollon, Switzerland,
1995}, edited by B.~Guiderdoni and J.~Tran Thanh V\^an (Editions
Fronti\`eres, Gif--sur--Yvette, 1995); R.~Brandenberger,
B.~Carter, A.~C.~Davis, and M.~Trodden, Phys. Rev. D {\bf 54},
6059 (1996); B.~Carter, A.~C.~Davis, {\it ibid} {\bf 61}, 123501
(2000).

\bibitem{rdp}  A.~Riazuelo, N.~Deruelle, and P.~Peter, Phys. Rev. D {\bf 61},
123504 (2000).

\bibitem{CMB} C.~L.~Bennett et al., Ap.J.Suppl. {\bf 148}, 97 (2003);
 D.~N.~Spergel et al., Ap.J.Suppl. {\bf 148}, 175 (2003).

\bibitem{vanessa} P. Peter, M. E. X. Guimar\~aes and V. C.
Andrade, Phys. Rev. D {\bf 67}, 123509 (2003).

\bibitem{louis} L. Witten, in {\em Gravitation}, ed. L. Witten (Wiley, New York, 1962).

\bibitem{melvin}M. A. Melvin,  Phys. Lett. {\bf 8}, 64 (1964).

\bibitem{chandra} S. Chandrasekhar, in {\em ``Mathematical Theory
of the Black Holes"} (Claredon Press Oxford, 1983).

\bibitem{damour} Th. Damour, Astrophys. Space Sci. {\bf 283}, 445
(2003).

\bibitem{lee}T. H. Lee, Mod. Phys. Lett. A {\bf 19}, 2929 (2004).

\end{thebibliography}
\end{document}